\documentclass[aps,prl,superscriptaddress,twocolumn,showpacs]{revtex4}

\usepackage{graphicx}
\usepackage{times}
\usepackage{amsmath}
\usepackage{amsthm}
\usepackage{amssymb}
\usepackage{amsbsy}
\newcommand{\abs}[1]{\vert #1 \vert}
\newcommand{\bra}[1]{\langle #1 \vert}
\newcommand{\ket}[1]{\vert #1 \rangle}

\begin{document}
\title{Implementing arbitrary phase gates with Ising anyons}
\author{Parsa Bonderson}
\affiliation{Microsoft Research, Station Q, Elings Hall, University of
California, Santa Barbara, California 93106, USA}

\author{David J.~Clarke}
\affiliation{Department of Physics and Astronomy,
University of California, Riverside, California 92521, USA}

\author{Chetan Nayak}
\affiliation{Microsoft Research, Station Q, Elings Hall, University of
California, Santa Barbara, CA 93106, USA} \affiliation{Department of
Physics, University of California, Santa Barbara, California 93106, USA}

\author{Kirill Shtengel}
\affiliation{Department of Physics and Astronomy, University of
California, Riverside, California 92521, USA}

\begin{abstract}
Ising-type non-Abelian anyons are likely to occur in a number of physical
systems, including quantum Hall systems, where recent experiments support
their existence. In general, non-Abelian anyons may be utilized to provide
a topologically error-protected medium for quantum information processing.
However, the topologically protected operations that may be obtained by
braiding and measuring topological charge of Ising anyons are precisely
the Clifford gates, which are not computationally universal. The Clifford
gate set can be made universal by supplementing it with single-qubit
$\pi/8$-phase gates. We propose a method of implementing arbitrary single-qubit
phase gates for Ising anyons by running a current of anyons with
interfering paths around computational anyons. While the resulting phase
gates are not topologically protected, they can be combined with ``magic
state distillation'' to provide error-corrected $\pi/8$-phase gates with
a remarkably high threshold.
\end{abstract}

\maketitle

Non-Abelian anyons -- quasiparticles with exotic exchange statistics
described by multidimensional representations of the braid
group~\cite{Leinaas77,Goldin85,Fredenhagen89,Froehlich90} -- can provide
naturally fault-tolerant platforms for quantum computation. The non-local
state space of such anyons can be used to encode qubits that are
impervious to local perturbations. Topologically protected computational
gates may be implemented by braiding the
anyons~\cite{Kitaev03,Preskill98,Freedman98,Freedman03b} or by measuring
their topological charge~\cite{Bonderson08a,Bonderson08b}.

Ising-type anyons~\footnote{We use the term ``Ising-type'' to include
anyons which only differ from Ising by Abelian factors, e.g. can be
obtained from Ising through products or cosets with U$(1)$ sectors.}
currently appear to be the most likely platform on
which topological quantum computation will be actualized. They
are expected to occur in a number of systems, including second Landau
level quantum Hall
states~\cite{Moore91,Blok92,Lee07,Levin07,Bonderson07d}, $p_x +i p_y$
superconductors~\cite{Read00}, lattice models~\cite{Kitaev06a},
topological insulator-superconductor interfaces~\cite{Fu08}, and any
generic 2D system with Majorana fermions~\cite{Sau09}. The existence
of such non-Abelian anyons in the $\nu=5/2$ quantum Hall state is
supported by recent experiments~\cite{Radu08,Willett09a,Bishara09}.

The braiding transformations of Ising anyons are given by the spinor
representations of SO$(2n)$~\cite{Nayak96c}. The set of gates that
may be obtained through braiding and/or topological charge measurement of
Ising anyons is encoding-dependent, but never computationally universal.
For the standard qubit encoding (i.e. one qubit in four anyons), the
computational gates obtained via braiding or measurement of anyon pairs
are the single-qubit Clifford gates. These gates can be generated by
the Hadamard and $\pi /4$-phase gates,
\begin{equation}
H=\frac{1}{\sqrt{2}} \left[
\begin{array}{rr}
1  &  1 \\
1  & -1
\end{array}
\right]
\quad \text{and} \quad
R\left( \pi / 2 \right) =\left[
\begin{array}{rr}
1  &  0 \\
0  &  i
\end{array}
\right]
,
\end{equation}
where $R\left( \theta \right) = \text{diag} \left[ 1,e^{i \theta} \right]$
is called the ``$\theta /2$-phase gate.''

The controlled-NOT gate
\begin{equation}
\text{CNOT} = \left[
\begin{array}{rrrr}
1  &  0  &  0  &  0 \\
0  &  1  &  0  &  0 \\
0  &  0  &  0  &  1 \\
0  &  0  &  1  &  0
\end{array}
\right]
\end{equation}
may be implemented by allowing the use of non-demolitional measurements of the collective topological charge of four anyons~\cite{Bravyi05,Bravyi06,BondersonWIP}. Adding this generates the full set of Clifford
gates, which can be efficiently simulated
on a classical computer, but becomes universal when supplemented
with a single-qubit $\pi/8$-phase gate~\cite{Boykin99}.

One way to obtain $\pi/8$-phase gates (as well as CNOT gates) is through dynamical topology change of the
system~\cite{Bravyi00-unpublished,Freedman06a}. However, this requires
complicated physical manipulations of the system which are (at best)
currently infeasible, such as switching between planar and non-planar
geometries.

Alternatively, if one can implement ideal (e.g. topologically protected) Clifford gates,
then they can be used to perform ``magic state distillation''~\cite{Bravyi05,Bravyi06}
to produce error-corrected $\pi/8$-phase gates from noisy ones.
This purification protocol (which has poly-log overhead) consumes several copies
of a magic state, e.g. $\left| A_{\pi/4} \right\rangle = \frac{1}{\sqrt{2}} ( \left| 0 \right\rangle + e^{i \pi/4} \left| 1 \right\rangle )$,
and outputs a single qubit with higher polarization along a magic direction.
Once a sufficiently pure magic state is produced, it may then be consumed to generate a $\pi/8$-phase gate.
This protocol permits a remarkably high error threshold of over $0.14$ for the noisy gates, as compared to the ``high'' threshold of $10^{-3}$ for postselected quantum computation~\cite{Aliferis07}.
Hence, it is important to devise practical methods of
generating the $\pi/8$-phase gate within this error threshold for systems with Ising anyons.

A simple proposal for this is to move bulk quasiparticles close
enough to each other to let the microscopic physics split the energy
degeneracy of the fusion channels encoding a qubit. The resulting time
evolution can produce arbitrary phase gates,
albeit unprotected ones in need of error-correction (e.g. by magic state
distillation). However, the energy splitting caused by bringing two
quasiparticles together oscillates rapidly with their
separation~\cite{Baraban09,Cheng09}, so small errors in the
quasiparticles' spatial separation will translate into large errors in the
phase. Thus, this approach appears unlikely to be able to meet even the
generous error threshold of magic state distillation.

In this letter, we propose a method of implementing arbitrary phase gates
for systems with Ising-type anyons that aims to be more practical and to
achieve a manageable error rate. This method involves a device consisting
effectively of a beam-splitter or tunneling junction that is used to run a
current of anyons through interfering paths around computational anyons.
We first analyze the effect of such a device using a semiclassical picture
of the anyonic current applicable to general Ising systems. Subsequently,
we perform a more detailed analysis (including error estimates) for
Ising-type systems in which the anyonic current is provided
by edge modes described by conformal field theory.

\begin{figure}[t]
\includegraphics[width=0.6\columnwidth]{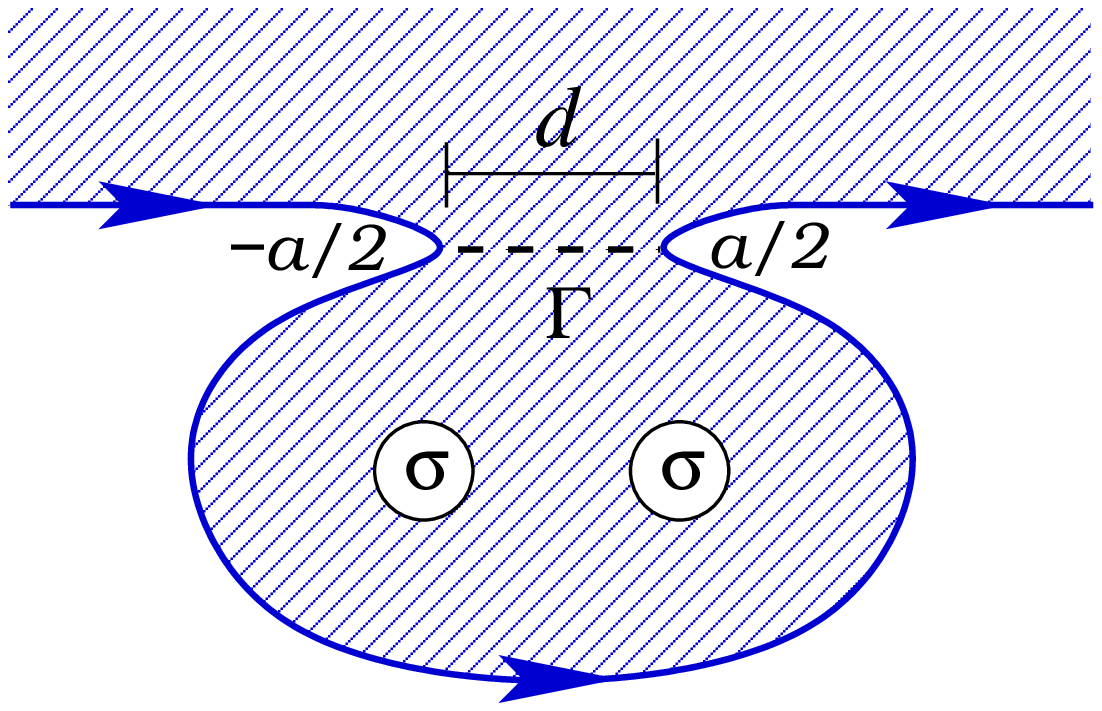}
   \caption{An implementation of the phase gate device in Ising-type quantum Hall states.
   A section of the Hall fluid (hatched region) is formed into a sack like
   enclosure around two $\sigma$ anyons encoding a qubit.
   The edge current (arrowed lines) tunnels quasiparticles across the constriction
   with strength $\Gamma$, inducing a phase gate on the qubit.}
\label{fig:gate}
\end{figure}

For the purpose of constructing the phase gate, we consider a topological
qubit encoded in a pair of anyons carrying Ising topological charge
$\sigma$~\footnote{Strictly speaking, more than two anyons are needed to
comprise a topological qubit and we will assume the standard one qubit in
four anyons encoding, but the operation we perform only requires
manipulations involving two of these anyons.}. The two possible fusion
channels $I$ and $\psi$ correspond to the qubit basis states $\left| 0
\right\rangle$ and $\left| 1 \right\rangle$. In the quantum Hall context,
the anyons comprising the topological qubits may be localized using
quantum antidots, each carrying a topological charge of $\sigma$ (i.e. an
odd number of $e/4$ quasiparticles), as proposed in~\cite{DasSarma05}. The
two anyons comprising a qubit are placed in a ``sack'' geometry as shown
in Fig.~\ref{fig:gate}. This geometry was proposed for the detection of
non-Abelian statistics in~\cite{Hou06}, where it was described as a
``wormhole.'' The sack may be created by deforming the edge of the quantum
Hall droplet. The current flowing around the edge can tunnel from $-a/2$
to $a/2$ (so the sack has perimeter length $a$) with a strength determined
by the distance $d$ across the constriction. In the weak-tunneling, low
temperature, low voltage limit, the quasiparticles with the most relevant
tunneling operators will dominate the tunneling current. For Ising-type
quantum Hall states, this will generally include but not be restricted to
the fundamental quasiholes that carry charge $\sigma$. However,
quasiparticles that do not carry $\sigma$ will have no effect on the
topological qubit here, so we will neglect them in our analysis. As we
will see, the interference between the possible trajectories from left to
right enacts a non-trivial transformation on the qubit.

In other possible physical realizations of Ising anyons, the $\sigma$
anyons comprising a qubit may need to be pinned by other means, e.g. in a
chiral $p$-wave superconductor, a hole may be bored through the sample
where flux can be trapped. It may be also be easier in some realizations
to construct interfering paths for a beam of bulk quasiparticles, rather
than to rely on edge quasiparticles. With this situation in mind, we now
compute semiclassically the effect of a beam of $\sigma$ quasiparticles
incident from the left. This calculation will also capture some of the
features of the more involved edge theory calculation, relevant to the
quantum Hall setting. For ease of comparison with Fig.~\ref{fig:gate}, we
use terminology appropriate to that picture. We assume that at the
tunneling junction in Fig.~\ref{fig:gate} a $\sigma$ quasiparticle can
tunnel with amplitude $\mathcal{T}$ and will continue along the edge with
amplitude $\mathcal{R}$.

We can treat the motion of the anyons semiclassically and analyze the
effect that sending them through the device has on the qubit. This effect results from
the braiding statistics, which contributes a factor of $+1$ or $-1$ when a
$\sigma$ anyon travels one full circuit around a region containing
topological charge $I$ or $\psi$, respectively. This
non-Abelian contribution is in addition to the Abelian phase $\alpha$
acquired when a $\sigma$ anyon travels once around the device loop
counterclockwise. This phase $\alpha$ contains the Abelian statistical
angle, the Aharanov-Bohm phase, and possibly other terms, depending on the
specific realization of the device.

The resulting transformation to the qubit when one $\sigma$
anyon has passed through the device is
\begin{eqnarray}
U &=& \mathcal{T} e^{-i\alpha} \boldsymbol{\sigma}_z   + { \left| \mathcal{R} \right|^{2} }
\sum_{n=0}^{\infty} \left( -\mathcal{T}^{\ast} e^{i \alpha} \boldsymbol{\sigma}_z \right)^{n}
\nonumber \\
&=& \left[
\begin{array}{cc}
\frac{ 1 + \mathcal{T} e^{-i \alpha} }{ 1 + \mathcal{T}^{\ast} e^{i \alpha} }     &   0   \\
0  &   \frac{ 1 - \mathcal{T} e^{-i \alpha} }{ 1 - \mathcal{T}^{\ast} e^{i \alpha} },
\end{array}
\right]
\end{eqnarray}
where $\boldsymbol{\sigma}_z$ accounts for the non-Abelian braiding
statistics. Here the first term results from direct tunneling across the
constriction, and the remaining terms describe the effect of the $\sigma$
quasiparticle passing around the edge of the sack one or many times. This
does not transfer topological charge to the qubit, so the matrix $U$ is
diagonal and unitary. However, braiding a $\sigma$ quasiparticle from the
beam around the computational $\sigma$ anyons is topologically equivalent
to processes that transfer topological charge $\psi$ between the
computational pair. These are the same processes that would cause energy
splitting between the otherwise degenerate fusion channels of the qubit
when its $\sigma$ quasiparticles are brought close
together~\cite{Bonderson09b}. Hence, the net effect of passing a $\sigma$
through the device is similar to that of splitting the energy, i.e. to
produce a relative phase between these channels. Up to an overall phase,
$U = R \left( \theta \right)$ where
\begin{equation}
\label{eq:theta_ideal}
\theta = 2 \arctan \left[ \frac{ 2 \left| \mathcal{T} \right| \sin \gamma }
{ 1 - \left| \mathcal{T} \right|^{2} } \right]
,
\end{equation}
and $\gamma = \alpha-\arg\left\{\mathcal{T} \right\}$. For $\left|
\mathcal{T} \right| \ll 1$, this gives $\theta \simeq 4 \left| \mathcal{T}
\right| \sin \gamma $. The phase gate generated using this device may be
controlled by sending multiple $\sigma$ quasiparticles through the system,
or by adjusting the experimental variables $\mathcal{T}$ and $\alpha$.

For Ising-type systems that support an anyonic edge current, such as those in Refs.~\cite{Moore91,Blok92,Lee07,Levin07,Bonderson07d,Read00,Kitaev06a,Fu08,Sau09}, we should go beyond this semiclassical calculation and analyze
the quasiparticle tunneling and interference using the proper edge theory. The combined edge and qubit system is
described by the Hamiltonian
\begin{equation}
H=H_E \otimes \openone + H_{\mathrm{tun}}(t) \otimes \boldsymbol{\sigma}_z,
\end{equation}
where $H_E$ is the Hamiltonian describing the unperturbed edge and
$H_{\mathrm{tun}}$ describes tunneling of $\sigma$ quasiparticles across
the constriction. As before, the $\boldsymbol{\sigma}_z$ represents the braiding
statistics of the edge $\sigma$ with the qubit, picking up a minus sign
each time the $\sigma$ braids around the $\psi$ charge. The strength of
the tunneling Hamiltonian can be adjusted by changing the separation
distance $d$ across the sack constriction. We represent the density
matrix of the combined system by $\chi$ and the qubit's density matrix is
obtained from this by tracing out the edge $\rho = \text{Tr}_{E} \chi$.

Solving the interaction picture Schr\"{o}dinger equation
\begin{equation}
\label{eq_density}
i \frac{\mathrm{d}\widetilde{\chi} (t) }{\mathrm{d}t}
= [\widetilde{H}_\mathrm{tun}(t)\otimes\boldsymbol{\sigma}_z,\widetilde{\chi}(t)],
\end{equation}
where $\widetilde{A}(t) = e^{i H_E (t-t_0)} A(t) e^{-i H_E (t-t_0) } $, with the
assumption that the edge and qubit are unentangled at time $t=t_0$, we find that
\begin{equation}
\rho (t) = \left[
\begin{array}{cc}
\rho_{00} (t_0)                        &     e^{-\varsigma^2/2} e^{-i \theta} \rho_{01} (t_0) \\
e^{-\varsigma^2/2}e^{i \theta} \rho_{10} (t_0)     &     \rho_{11} (t_0)
\end{array}
\right],
\end{equation}
where $\theta$ and $\varsigma^2$ are real-valued time dependent
quantities, and $\varsigma^2 \geq 0$. The diagonal elements of the qubit
density matrix are unaltered from their initial state, as the Hamiltonian
commutes with $\openone\otimes \boldsymbol{\sigma}_z$. The sack geometry
will therefore implement a phase gate $R(\theta)$ with phase-damping noise
parameterized by $\varsigma^2$.

Applying a Hadamard gate and then this noisy phase gate to an initial state $\ket{0}$
creates a magic state $\ket{A_{\pi/4}}$ with error
\begin{equation}
\epsilon=1-\bra{A_{\pi/4}}\rho\ket{A_{\pi/4}}
=\frac{1}{2}\left[1-e^{-\varsigma^2/2}\cos\!\left(\theta-\frac{\pi}{4}\right)\right].
\end{equation}
If $\epsilon< 0.14$, then this can be used with magic state distillation to
generate an error-corrected $\pi/8$-phase gate~\cite{Bravyi05,Bravyi06}.

Computing the values of $\theta$ and $\varsigma^2$ to second order in the
tunneling Hamiltonian, we have
\begin{eqnarray}
\label{eq:theta_FQH}
\!\! \theta &\simeq&  2\int_{t_0}^{t}~\mathrm{d}t'
\left\langle \widetilde{H}_\mathrm{tun}(t') \right\rangle, \\
\label{eq:p_FQH}
\!\! \varsigma^2 &\simeq& -\theta^{2} + 4 \int_{t_0}^{t} \mathrm{d}t_1
\int_{t_0}^{t}~\mathrm{d}t_2 \left\langle \widetilde{H}_\mathrm{tun}(t_1)
\widetilde{H}_\mathrm{tun}(t_2) \right\rangle.
\end{eqnarray}
We note that $\varsigma^2$ takes the form of a variance in the phase.

To compute concrete values of $\theta$ and $\varsigma^2$ for the most physically relevant example,
we turn to the field theoretic description of the edge of a Moore-Read (MR) quantum Hall state~\cite{Moore91}.
The Lagrangian for the unperturbed edge is~\cite{Fendley07a}
\begin{equation}
\label{lagrange}
\mathcal{L}_E = \frac{1}{2\pi}\partial_x\varphi(\partial_t+v_c\partial_x)\varphi
+i\psi(\partial_t+v_n\partial_x )\psi ,
\end{equation}
where the charged and neutral sectors are respectively
described by the chiral boson ($\varphi$) and fermion ($\psi$) modes, with velocities $v_{c}$ and $v_{n}$.
The operator that tunnels $\sigma$ quasiparticles with charge $e^{*}=e/4$ across the constriction is
\begin{equation}
H_{\mathrm{tun}} = \Gamma e^{-i\alpha} \, \sigma \! \left(\mbox{$\frac{a}{2}$}\right) \sigma \!
\left( -\mbox{$\frac{a}{2}$} \right) e^{ i \frac{\varphi (a/2)}{\sqrt{8}}}
e^{- i \frac{\varphi(-a/2)}{\sqrt{8}}} + h.c.,
\end{equation}
where $\alpha$ includes the Aharanov-Bohm phase ($e^{*}BA$) acquired in traveling
around the sack as well as any Abelian braiding statistics factors.

Assuming the edge was initially in thermal equilibrium at temperature $T$,
i.e. $\chi(t_0) = \frac{ e^{-H_{E}/T}}{ \text{Tr}_{E} \left[ e^{-H_{E}/T}
\right] } \otimes \rho(t_0)$, we find
\begin{equation}
\label{eq:gatefreq}
\left\langle \widetilde{H}_\mathrm{tun}(t) \right\rangle \! =2 \! \left(\frac{\lambda\pi T/ v_c}
{\sinh{\frac{\pi T a}{v_c}}}\right)^{ g_c } \!\!\!
\left(\frac{\lambda\pi T/ v_n}{\sinh{\frac{\pi T a}{ v_n}}}\right)^{ g_n }
\!\!\! \abs{\Gamma}\sin\gamma.
\end{equation}
Here $\lambda$ is a short range cutoff, $g_c=1/8$ and $g_n=1/8$ are the
scaling exponents of the charge and neutral modes, respectively, and
$\gamma = \alpha - \arg \left\{ \Gamma \right\} + \pi/2$.

From Eqs.~(\ref{eq:theta_FQH},\ref{eq:gatefreq}), we see that there are
several experimental parameters which may be used to control the phase
$\theta$ generated using the sack geometry. In particular, we envision $d$
and the area $A$ enclosed in the sack as the primary physical quantities
to adjust, since these provide a practical means of tuning $\abs{\Gamma }$
and $\gamma$, respectively, while keeping the other quantities essentially
constant. With a properly designed geometry, these quantities can be
adjusted sufficiently while causing only negligible changes to $a$. In contrast to
the tunneling amplitude of neutral $\psi$ excitations, which oscillates rapidly with distance
\cite{Baraban09,Cheng09} (and can be understood as Friedel oscillations
in a composite fermion picture), the tunneling amplitude of $\sigma$
quasiparticles does not oscillate and decays as $\Gamma \sim e^{ -\left(e^{\ast} d / 2 e \ell_{B} \right)^{2}}$ for $d \gg \ell_B$, where $\ell_B$ is the magnetic length~\cite{Bishara09,Chen09}.

There are several ways to adjust the phase $\gamma$ for quantum Hall systems. One practical method is to alter the
total area enclosed in the sack by using a side gate. This leads to a change in the flux enclosed in the two
interfering current paths, and thus a change in the Aharanov-Bohm phase
included in $\gamma$. Another method for changing $\gamma$ is by applying a current along the edge of the system.
This may be implemented via a voltage difference between the edge that forms the sack structure
and the edge on the other side of the electron gas. Driving this
current populates or depopulates charge on the edge of
the electron gas, and hence changes the area as a side gate would.

Let us hold fixed all the experimental parameters except the tunneling amplitude, which we vary as $\Gamma(t) = \Gamma_{0} f(t)$, for $f(t)$ a general (real, non-negative) signal profile with characteristic ``duration'' time scale $\tau \equiv \int_{-\infty}^{\infty} \mathrm{d}t f(t)$. This gives $\theta \simeq \omega \tau$, where
\begin{eqnarray}
\omega &\equiv& 2 \langle\widetilde{H}_\mathrm{tun}(t) \rangle /f(t) \\
&=& 4 \! \left(\frac{\lambda\pi T/ v_c}
{\sinh{\frac{\pi T a}{v_c}}}\right)^{ g_c } \!\!\!
\left(\frac{\lambda\pi T/ v_n}{\sinh{\frac{\pi T a}{ v_n}}}\right)^{ g_n }
\! \abs{\Gamma_0} \sin\gamma
,
\end{eqnarray}
and
\begin{eqnarray}
\varsigma^{2} &\simeq& \frac{\omega^{2}}{\sin^{2} \gamma} \int_{-\infty}^{\infty} \mathrm{d}t \, \eta(t) F(t), \\
\eta(t) &=& \frac{1}{2}\left(\Upsilon_{c}^{g_{c}} \Upsilon_{n}^{g_{n} - \frac{1}{4}} - \Upsilon_{c}^{-g_{c}} \Upsilon_{n}^{-g_{n}}\right) \sqrt{ \frac{1+\Upsilon_{n}^{\frac{1}{2}}}{2}} \notag \\
&&+ \left( \Upsilon_{c}^{-g_{c}} \Upsilon_{n}^{-g_{n}} \sqrt{ \frac{1+\Upsilon_{n}^{\frac{1}{2}}}{2} } -1 \right) \sin^{2} \gamma , \\
\!\!\!\! \Upsilon_{c,n}(t) &=& 1- \frac{\sinh^2\left(\frac{ \pi T a}{ v_{c,n}}\right)} {\sinh^2\left(\pi T t - i \delta\right)},
\end{eqnarray}
where $F(t) \equiv \int_{-\infty}^{\infty} \mathrm{d}t^{\prime} f(t^{\prime}) f(t^{\prime} - t)$.
We note that $\eta(t)\rightarrow 0$ exponentially for long times with a decay rate
proportional to the temperature. We generally have the bound
\begin{equation}
\varsigma^{2} \lesssim \frac{\omega^{2} }{ \sin^{2}\gamma}\frac{\sinh \left(\frac{ \pi T a}{ v_{n}}\right)}{ \pi T } \kappa \tau
\end{equation}
where $\kappa$ is a dimensionless function of $g_{c}$, $g_{n}$, and $\frac{\sinh \left(\frac{ \pi T a}{ v_{c}}\right)}{\sinh \left(\frac{ \pi T a}{ v_{n}}\right)}$. When $\frac{a}{v_{c,n}} \ll \frac{1}{\pi T}$ (i.e. $a$ is much shorter than the thermal coherence length), this becomes a temperature independent bound $\varsigma^{2} \lesssim \frac{\omega^{2} a/ v_n }{ \sin^{2}\gamma} \kappa \tau$ with $\kappa$ now depending only on $g_{c}$, $g_{n}$, and $v_{n}/v_{c}$. Using $\omega \simeq \theta / \tau$ in these expressions, we see that it is favorable to increase the duration $\tau$ (e.g. by using weaker tunneling) used to enact a particular phase gate, since the bound decreases as $1/\tau$. However, one must obviously balance this with the need to keep time scales much shorter than the qubits' coherence time.

\begin{figure}[t]
\includegraphics[width=.8\columnwidth]{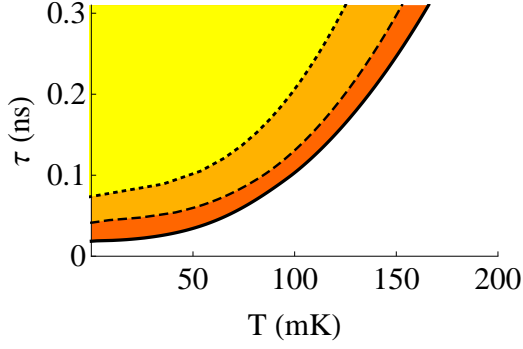}
  \caption{Error-correction threshold curves for implementing a $\pi/8$-phase gate using the sack geometry in a Moore-Read state, when $\gamma = \pi/2$ (solid), $\pi/8$ (dashed), and $\pi/16$ (dotted). Magic state distillation is applicable in the shaded region above these curves (which can thus be viewed as indicating the minimum required gate duration as a function of temperature).}
\label{fig:error}
\end{figure}

To demonstrate that the $\pi/8$-phase gate can (at least in principle) be implemented with sufficiently low error using this device, we compute $\varsigma$ resulting for $\theta = \pi/4$ using a sack of length $a=1\mu$m, a rectangular pulse of duration $\tau$, i.e. 
\begin{equation}
f(t) = \left\{
\begin{array}{ccc}
1  &  \text{ for } & \abs{t} < \frac{\tau}{2} \\
0  &  \text{ for } & \abs{t} > \frac{\tau}{2}
\end{array}
\right.
\end{equation}
\begin{equation}
F(t) = \left\{
\begin{array}{ccc}
\tau - \abs{t}  &  \text{ for } & \abs{t} < \tau \\
0  &  \text{ for } & \abs{t} > \tau
\end{array}
\right.
,
\end{equation}
and velocities \mbox{$v_c=10^5$ m/s} and \mbox{$v_n=10^4$ m/s} estimated for the $\nu=5/2$ state from numerical studies~\cite{Wan08a,Hu09}. In Fig.~\ref{fig:error}, we display the resulting region of parameter space (for different values of $\gamma$) in which the error is below the threshold $\epsilon< 0.14$ for magic state distillation. The threshold curves move up as $\gamma$ is varied away from $\pi/2$, and will diverge as $\gamma \rightarrow 0$ or $\pi$. However, this divergence is evidently not problematic unless $\gamma$ is rather close to the singular points.

It is straightforward to repeat the preceding edge theory analysis for other Ising-type systems with a conformal edge theory.
The results are again given by the preceding equations, but with different values of $g_c$, $g_n$,
$e^{\ast}$, and $\alpha$. The values of these quantities for Ising-type quantum Hall candidates for all the observed second Landau level plateaus, such as the MR state~\cite{Moore91}, SU$(2)_2$ NAF state~\cite{Blok92}, the anti-Pfaffian state~\cite{Lee07,Levin07}, and the Bonderson-Slingerland hierarchy states~\cite{Bonderson07d} built on any of these Ising-type states, can be found in~\cite{Bishara09}. For systems with chargeless Ising edges~\cite{Read00,Kitaev06a,Fu08,Sau09}, one has $g_c = e^{\ast} =0$ and $g_n = 1/8$.

We also note that we can use our device to generate the two-qubit gate: $\text{diag} \left[ 1,e^{i \theta},e^{i \theta},1 \right]$ by putting two pairs of $\sigma$ anyons, each pair corresponding to a separate encoded qubit, into the sack. This is an entangling gate if $\theta \neq n \pi$ for $n \in \mathbb{Z}$, and in particular is a Clifford gate when $\theta = \pm \pi/2$. Similarly, this device can be used to generate multi-qubit gates.

In addition to offering a correctable error rate, the phase gate
implementation described herein offers several advantages that increase
its practicality. This device may be utilized in a manner compatible with proposals for ``measurement-only'' topological quantum computation~\cite{Bonderson08a,Bonderson08b}. Specifically, the computational anyons
may remain stationary while only the edge of the system is manipulated, thus circumventing the need for fine control
over the motion of bulk quasiparticles. As this device would only require the use of established techniques for deforming the edge using top and side
gates~\cite{Willett09a}, it provides the first realistic proposal for achieving universal quantum computation using Ising anyons.

\begin{acknowledgments}
We thank S.~Simon for useful discussions. DC and KS
acknowledge the support and hospitality of Microsoft Station Q. DC, CN and
KS are supported in part by the DARPA-QuEST program. KS is supported in
part by the NSF under grant DMR-0748925.\end{acknowledgments}

\end{document}